\documentclass[aps,prd,preprintnumbers,showpacs,nofootinbib,twocolumn]{revtex4}
\usepackage{graphicx}
\usepackage{bm}
\usepackage{amsmath}
\usepackage{amssymb}
\usepackage{amsthm}
\input{epsf}

\newcommand{\lsim}   {\mathrel{\mathop{\kern 0pt \rlap
  {\raise.2ex\hbox{$<$}}}
  \lower.9ex\hbox{\kern-.190em $\sim$}}}
\newcommand{\gsim}   {\mathrel{\mathop{\kern 0pt \rlap
  {\raise.2ex\hbox{$>$}}}
  \lower.9ex\hbox{\kern-.190em $\sim$}}}

\begin{document}

\title{Cosmological magnetic fields from inflation 
in extended electromagnetism}

\author{Jose Beltr\'an Jim\'enez}
\author{Antonio L. Maroto}

\affiliation{Departamento de F\'isica Te\'orica, Universidad Complutense 
de Madrid, 28040, Madrid, Spain.}


\date{\today}

\begin{abstract}
In this work we consider an extended electromagnetic theory in which
the scalar  state which is usually eliminated by means
of the Lorenz condition is allowed to propagate. This state has
been shown to generate a small    
cosmological constant in the context 
of standard inflationary cosmology. Here we show that  
the usual Lorenz gauge-breaking term now plays the role of an effective
electromagnetic current. Such a current is generated during inflation 
from quantum fluctuations and gives rise to a stochastic effective 
charge density
distribution. Due to the high electric conductivity of the cosmic
plasma after inflation, the electric charge density generates currents which 
give rise
to both vorticity and magnetic fields on sub-Hubble scales. Present
upper limits on vorticity coming from temperature anisotropies of the CMB 
are translated into {\it lower limits} on the present value of cosmic magnetic
fields.  We find that,  for
a nearly scale invariant vorticity spectrum,
 magnetic fields
$B_{\lambda}> 10^{-12}$ G are typically generated with 
coherence lengths ranging from  sub-galactic
scales  up to the present Hubble radius.  
 Those fields could
 act as seeds for a galactic
dynamo or even account for observations  just by collapse
and differential rotation of the protogalactic cloud.

\end{abstract}

\maketitle


Traditionally it has been argued  
that due to the electric neutrality of the universe on 
large scales, the only relevant interaction in cosmology 
should be gravitation. However, the behaviour of electromagnetic fields  
on astrophysical and cosmological scales  
is still far from clear, the most evident example being the 
unknown origin of  magnetic fields observed in galaxies 
and galaxy clusters. 

Magnetic fields with  
 large coherence lengths (around 10 kpc or even larger), 
and strengths around $10^{-6}$ G 
\cite{Widrow} have been measured in galaxies of all types and 
in galaxy clusters located at very different redshifts. 
Also,  recent works \cite{extragalactic} show
evidence for the existence of strong extragalactic magnetic fields 
above $3\times 10^{-16}$ G with coherence lengths much larger than 
the cluster scales
and eventually reaching the present Hubble radius. 

Two different types of scenarios
have been considered for the generation of such fields. 
On one hand, the primordial field hypothesis, i.e. 
the existence of  relic magnetic fields from the early universe 
 with comoving strengths around $10^{-10}-10^{-12}$ G 
which permeated the protogalactic medium
and were amplified to the present values 
by collapse and differential rotation. On the other hand, 
we have the dynamo
mechanism, in which much
weaker fields, around $10^{-19}$ G \cite{Brandenburg},  
could have been amplified by the galactic rotation. 
However, it is known that the latter scenario has certain limitations 
since  
the timescales for 
dynamo amplification may be too long to explain the observed fields in 
young objects \cite{young}. 

Both scenarios  require 
preexisting seed fields to be amplified and  
proposals for their generation include: 
astrophysical mechanisms \cite{Harrison}, 
production during inflation \cite{Turner}, in phase transitions \cite{QLS},
and others \cite{pert}. Although some of those 
mechanisms could seed a galactic dynamo, the generation of the 
stronger seeds required in the primordial field hypothesis is much more
problematic. In any case, according to 
\cite{extragalactic}, astrophysical processes, 
generation in phase transitions or
during recombination  
could not explain the claimed
extragalactic detection, and, in principle, only production 
during inflation could account for observations. 
Generation of
magnetic fields during inflation requires the breaking of the  
conformal triviality of standard electromagnetism in a Robertson-Walker
background. For that reason, modified electromagnetic theories, including
non-minimal curvature couplings  or couplings to 
extra fields such as inflaton or dilaton 
have been  studied in the literature 
\cite{Turner}. 

Recently,   the possibility of producing a 
small cosmological constant during inflation 
in  the context of an extended  electromagnetic model 
has been considered in  \cite{EM1,EM2,EM3}.
 The proposed theory  involves a modification of 
the  non-transverse electromagnetic sector, which breaks
conformal triviality but respects
 the ordinary (transverse) 
photons dynamics. Unlike previous models, 
the modified equations remain linear,
without potential
terms, dimensional parameters or explicit curvature couplings. 
The aim of this work will be to explore the possibility that large-scale
cosmic magnetic fields
could be
generated in  this extended theory.

Let us start by writing the generalized electromagnetic 
action   
which includes, apart from the coupling to the conserved 
 current $J^\mu$, 
 a gauge-breaking term \cite{EM1,EM2}:
\begin{eqnarray}
S=\int d^4x
\sqrt{g}\left[-\frac{1}{4}F_{\mu\nu}F^{\mu\nu}+\frac{\xi}{2}
(\nabla_\mu A^\mu)^2+ A_\mu J^\mu\right].\label{action}
\end{eqnarray}
Because of the presence of the gauge-breaking term, this
action does not respect the invariance under arbitrary gauge 
transformation, but it still preserves a residual gauge
symmetry given by $A_\mu\rightarrow A_\mu+\partial_\mu\theta$
provided $\Box\theta=0$. 

The corresponding modified 
Maxwell equations  read:
\begin{eqnarray}
\nabla_\nu F^{\mu\nu}+\xi\nabla^\mu(\nabla_\nu A^\nu)=J^\mu.
\label{EMeq}
\end{eqnarray}
Taking the 4-divergence of these equations we obtain:
\begin{eqnarray} \label{box}
\Box(\nabla_\nu A^\nu)=0\label{minimal}
\end{eqnarray}
where we have used the fact that the electromagnetic current is
covariantly conserved.

Thus we see that due to the presence of the $\xi$-term, the free theory 
contains three propagating physical fields, which correspond
to the two ordinary transverse photons and a third  scalar
state related to $\nabla_\nu A^\nu$ (in principle, 
we should include a fourth polarization for $A_\mu$, however it can be 
seen to correspond to the pure gauge mode $\partial_\mu \theta$). Notice that in the ordinary 
approach to electrodynamics  \cite{Itzykson}, 
the same action (\ref{action})
is considered, but 
$\nabla_\nu A^\nu$ is imposed to be zero (Lorenz condition), so that 
we are left only with the two transverse polarizations. However 
in the modified approach we will follow, we  
allow this state to propagate. Despite the fact that the 
extended theory is not gauge invariant, the transverse photons 
dynamics is not affected and remains gauge invariant. This implies that 
ordinary QED phenomenology is recovered in Minkowski space-time.
On the other hand, the fact that the theory contains an additional
polarization could suggest the possibility that such a mode is a ghost
and the theory would be quantum-mechanically unstable. 
However, as shown in  \cite{EM2}, thanks to the residual gauge 
symmetry of the theory
it is possible to eliminate the ghost state, so that the new 
mode has positive norm (details on  theoretical 
and experimental aspects of the
theory can be found in \cite{EM1,EM2,EM3} and references therein). 

As seen from (\ref{minimal})
the new state is completely decoupled from the conserved currents,
although it is non-conformally coupled to gravity. This means
that this state cannot be excited from electromagnetic currents. 
However, it 
could  be produced from quantum fluctuations in a curved space-time, in
 a similar way as inflaton fluctuations during inflation.
 Moreover, due to the well-known fact that a massless scalar field
gets frozen on super-Hubble scales for a Robertson-Walker universe 
($ds^2=a^2(\eta)(d\eta^2-d\vec x^2)$), 
we get $\nabla_\nu A^\nu\sim$ const. on scales
larger than the Hubble radius, giving rise to a cosmological
constant-like term in the action (\ref{action}). This constant has been shown to agree
with observations provided inflation took place at the electroweak scale
\cite{EM1,EM2}.
On the other hand, for sub-Hubble
scales, we have that 
$\nabla_\nu A^\nu\sim a^{-1}e^{i(k\eta-\vec k\vec x)}$.

It is interesting to note that the $\xi$-term 
can be seen, at the equations of motion
level, as a conserved current acting as a source of the usual
Maxwell field. To see this, we can write
$-\xi\nabla^\mu(\nabla_\nu A^\nu)\equiv J_{\nabla\cdot A}^\mu$
which, according to (\ref{minimal}), satisfies the conservation
equation $\nabla_\mu J_{\nabla\cdot A}^\mu=0$ and we can express
(\ref{EMeq}) as:
\begin{eqnarray}
\nabla_\nu F^{\mu\nu}=J^\mu_T
\label{ModMax}
\end{eqnarray}
with $J^\mu_T=J^\mu+J^\mu_{\nabla\cdot A}$ and $\nabla_\mu
J^\mu_T=0$. Physically, this means that, while the new scalar mode
can only be excited gravitationally, once it is
produced it will generally behave as a source of electromagnetic
fields. Therefore,  the modified theory is described by 
ordinary Maxwell equations with an additional "external" current.

In the following we will study the phenomenological consequences of
the presence of this new effective current. We will show that it 
can be generated during inflation and we will compute its corresponding
power spectrum. This implies that the universe will acquire a non-vanishing
gaussian stochastic distribution of effective 
electric charge with zero mean but a
non-vanishing dispersion. 
This in vacuum,  could also be seen as the generation of a stochastic
background of longitudinal electric waves. 
Due to the high electric conductivity after inflation, 
an electrically charged universe 
has been shown to lead necessarily to 
the generation of vorticity and the presence of 
magnetic fields on cosmological scales \cite{Dolgov,FC}.
Notice that even though conductivity will be in general high 
after reheating also in ordinary electromagnetism, 
it is the presence of a non-vanishing effective
charge density the crucial ingredient leading to the generation of
cosmological magnetic fields. Finally, we will show that 
the existing upper limits on vorticity coming from CMB anisotropies 
impose a {\it lower limit} on the amplitude of  the 
produced magnetic fields.

The power spectrum of super-Hubble fluctuations of 
$\nabla_\nu A^\nu$ produced during 
an inflationary phase characterized by a slow-roll
parameter $\epsilon$,  can be written as \cite{EM1}:
\begin{eqnarray}
P_{\nabla A}(k)=\frac{9H_{k_0}^4}{16\pi^2}
\left(\frac{k}{k_0}\right)^{-4\epsilon}
\end{eqnarray}
where $H_{k_0}$ is the Hubble parameter when the 
$k_0$ mode left the horizon, and we have fixed $\xi=1/3$. 
As shown in \cite{EM1}, this value of  $\xi$ corresponds to  
canonical normalization of commutation relations for creation and
annihilation operators for states built out of the 
standard Bunch-Davies vacuum. Of course, in curved space-time 
it would be possible
to choose a different normalization condition, 
but as long as it is a natural choice, we do not expect deviations
from $\xi$ being of order unity and, thus, the results obtained 
in this work will remain essentially unchanged. 
The pivot point will be chosen as $k_0\simeq H_0$ with 
$H_0$ the Hubble parameter today. The corresponding field
variance will read:
\begin{eqnarray}
\langle (\nabla_\mu A^\mu)^2\rangle= \int_{k_c}^{H_0} \frac{dk}{k}
\frac{9H_{k_0}^4}{16\pi^2}\left(\frac{k}{k_0}\right)^{-4\epsilon}
\simeq \frac{9H_{k_0}^4}{64\pi^2\epsilon}\left(\frac{H_0}{k_c}\right)^{4\epsilon}
\nonumber \\
\end{eqnarray}
where $k_c$ is the infrared cutoff which is usually set
by the comoving Hubble radius at the beginning of inflation
(see \cite{Tanaka} and references therein for problems with infrared
divergences during inflation).
The above expression, for super-Hubble modes today,  
can be identified with the cosmological constant scale 
$M_\Lambda\simeq 2 \times 10^{-3}$ eV and, thus:
\begin{eqnarray}
M_\Lambda^4\simeq 
\frac{9H_{k_0}^4}{64\pi^2\epsilon}\left(\frac{H_0}{k_c}\right)^{4\epsilon}.
\label{CC}
\end{eqnarray}
Since $\epsilon$ is positive, we see that,
 in general, $H_{k0}\lsim M_\Lambda$. 
Notice that $\nabla_\nu A^\nu$ is constant on super-Hubble scales
 and starts
decaying as $1/a$ once the mode reenters the Hubble radius. Thus, today, 
a mode $k$ will have been suppressed by a 
factor  $a_{in}(k)$ (we are assuming that the 
scale factor today is $a_0=1$). This factor will be given by: 
$a_{in}(k)=\Omega_M H_0^2/k^2$ for 
modes entering the Hubble radius in the matter era, i.e. for 
$k<k_{eq}$ with $k_{eq}\simeq (14\, \mbox{Mpc})^{-1}\Omega_Mh^2$ 
the value of the mode which entered
at matter-radiation equality. For $k>k_{eq}$ we have 
$a_{in}(k)=\sqrt{2\Omega_M}(1+z_{eq})^{-1/2}H_0/k$.
It is then possible
to compute the corresponding power spectrum for the effective 
electric charge
density today $\rho_g^0=J_{\nabla\cdot A}^0=-\xi\partial_0(\nabla_\nu A^\nu)$.
Thus from:
\begin{eqnarray} 
\langle\rho(\vec k)\rho^*(\vec h)\rangle=
(2\pi)^3\delta(\vec k-\vec h)\rho^2(k)
\end{eqnarray}
we define 
$P_\rho(k)=\frac{k^3}{2\pi^2}\rho^2(k)$, which is given by:
\begin{eqnarray}
P_\rho(k)=\left\{
\begin{array}{cc}
0, & k<H_0\\
& \\
\frac{\Omega_M^2H_0^2 H_{k0}^4}{16\pi^2}
\left(\frac{k}{k_0}\right)^{-4\epsilon-2},& H_0<k<k_{eq}\\
&\\
\frac{2\Omega_M H_0^2 H_{k0}^4}{16\pi^2(1+z_{eq})}
\left(\frac{k}{k_0}\right)^{-4\epsilon},&  k>k_{eq}.
\end{array}\right. 
\label{chPE}
\end{eqnarray}
Therefore the corresponding charge variance will read:
 $\langle\rho^2\rangle=\int \frac{dk}{k}P_\rho(k)$.
Notice that for modes entering the Hubble radius in the
radiation era, the power spectrum is nearly scale
invariant. Also, due to the constancy of $\nabla_\nu A^\nu$ on super-Hubble 
scales, 
the effective charge density power spectrum is negligible on such scales, so that
we do not expect magnetic field nor vorticity generation on those
scales. Notice that, on sub-Hubble scales, the present amplitude of the 
longitudinal electric fields would be precisely $E_L\simeq \nabla_\nu A^\nu$.

For an observer moving with the cosmic plasma with four-velocity
$u^\mu$, it is possible to  decompose the Faraday tensor in its 
electric and magnetic parts  as: $F_{\mu\nu}=2E_{[\mu}u_{\nu ]}+\frac{\epsilon_{\mu\nu\rho\sigma}}
{\sqrt{g}}B^\rho u^\sigma$, where $E^\mu=F^{\mu\nu} u_\nu$ and
 $B^\mu=\epsilon^{\mu\nu\rho\sigma} /(2\sqrt{g})F_{\rho\sigma}u_\nu$.
In the infinite conductivity
limit, Ohm's law $J^\mu-u^\mu u_\nu J^\nu=\sigma F^{\mu\nu} u_\nu$
implies  $E^\mu=0$. Therefore, in that case the only contribution  
would come
from  the magnetic part.  Here, $J^\mu$ is the current generated in the 
plasma which is assumed neutral, i.e., $J_\mu u^\mu=0$. Thus, from (\ref{ModMax}), we get:
\begin{eqnarray}
F^{\mu\nu}_{\;\;\;\; ;\nu}u_{\mu}=
\frac{\epsilon^{\mu\nu\rho\sigma}}
{\sqrt{g}}B_\rho u_{\sigma \,;\nu}u_\mu=J^{\mu}_{\nabla\cdot A}u_{\mu}
\end{eqnarray}
that for  comoving observers in a Robertson-Walker metric imply
(see also \cite{FC}):
\begin{eqnarray}
\frac{1}{a^2}\vec \omega\cdot \vec B=\frac{\rho_g^0}{a^2}
\label{mag}
\end{eqnarray}
where $\vec v=d\vec x/d\eta$, $\vec\omega=\vec \nabla\times \vec v$ is the fluid 
vorticity, $\rho_g^0$ is the effective charge density today 
and the $\vec B$ components scale as $B_i\propto 1/a$ as can be
easily obtained from $\epsilon^{\mu\nu\rho\sigma}F_{\rho\sigma;\nu}=0$
to the lowest order in $v$. Thus, as commented before,  the 
presence of the non-vanishing cosmic effective charge density necessarily 
creates both magnetic field and vorticity. 
In the absence of sources,  
vorticity scales as $a^{3w-1}$ \cite{Durrer}, with $w$ the equation of 
state parameter of the dominant component. However, due to the 
presence of the 
effective current, we find that vorticity grows as $\vert\vec\omega\vert
\propto a$, from radiation era until present.

 Using (\ref{mag}), it is possible to estimate a 
lower limit on the present amplitude of the magnetic 
fields generated. Since we are not assuming any particular mechanism for 
the generation of the primordial magnetic and vorticity perturbations
in the early universe,  
we will consider them for simplicity as gaussian stochastic variables
such that:
\begin{eqnarray}
\langle B_i(\vec k)B_j^*(\vec h)\rangle&=&
\frac{(2\pi)^3}{2}P_{ij}\delta(\vec k-\vec h)B^2(k)\nonumber \\ 
\langle\omega_i(\vec k)\omega_j^*(\vec h)\rangle&=&
\frac{(2\pi)^3}{2}P_{ij}\delta(\vec k-\vec h)\omega^2(k)
\end{eqnarray}
with $B^2(k)=B k^n$, $\omega^2(k)=\Omega k^m$
and where $P_{ij}=\delta_{ij}-\hat k_i\hat k_j$ is 
introduced because of the 
transversality properties of $B_i$ and $\omega_i$.
The spectral indices $n$ and $m$ are in principle arbitrary. 
Notice that when the plasma conductivity becomes large after 
reheating, we expect constraints on the power spectra coming from (\ref{mag}). 
We will be interested in calculating 
the mean fluctuation of the magnetic field in a region of size $\lambda$ 
using a gaussian window function:
\begin{eqnarray}
B_\lambda^2&=&\frac{4\pi}{(2\pi)^3}\int \frac{dk}{k}k^3 B k^n
W^2(k\lambda)
\end{eqnarray}
where $W(k\lambda)=\exp(-k^2\lambda^2/2)$. Similar expressions can
be written for
$\omega_\lambda$ and $\rho_\lambda$.
  Thus from 
(\ref{mag}) it is possible to obtain \cite{FC,NucDurrer}: 
\begin{eqnarray}
\rho_\lambda^2&=&\frac{1}{(2\pi)^3}\frac{\Omega B}{2\pi^2}S(\lambda,n,m)
\end{eqnarray}
with:
\begin{eqnarray}
S(\lambda,n,m)&=&
\int_{H_0}^\infty \frac{dk}{k} k^3W^2(k\lambda)
\\
&\times&\int d^3p \vert \vec k-\vec p\vert^m p^n [1+(\widehat{\vec k-\vec p}\cdot
\widehat{\vec p})^2]\nonumber
\end{eqnarray}
where $H_0< p<k_{cB}$ and $\vert \vec k-\vec p\vert>H_0$. 
For the
upper cutoff of the magnetic power spectrum, 
we take a conservative value  
corresponding to the magnetic diffusion scale 
which is given by 
$k_{cB}\simeq 10^{11}$ Mpc$^{-1}$ \cite{CMBDurrer} 
(a more detailed analysis can be found in \cite{Barrow}).
Let us define:
\begin{eqnarray}
G(\lambda,n)=\int_{k_{min}}^{k_{max}} \frac{dk}{k}k^3  k^n
W^2(k\lambda)
\end{eqnarray}
where, as before, due to the vanishing of the charge density on 
super-Hubble scales, $k_{min}$ is typically given by the 
comoving Hubble horizon at the time the fluctuations  
are evaluated and $k_{max}=k_{cB}(\infty)$ in the magnetic (vorticity) 
cases respectively. Thus, we  finally obtain 
for the magnetic fluctuation on a scale $\lambda$:
\begin{eqnarray}
B_\lambda^2\simeq 
\frac{4\pi\rho_\lambda^2G(\lambda,n)G(\lambda,m)}
{\omega_\lambda^2S(\lambda,n,m)}.
\label{fund}
\end{eqnarray}

 \begin{figure}[ht!]
\begin{center}
{\epsfxsize=8.5cm\epsfbox{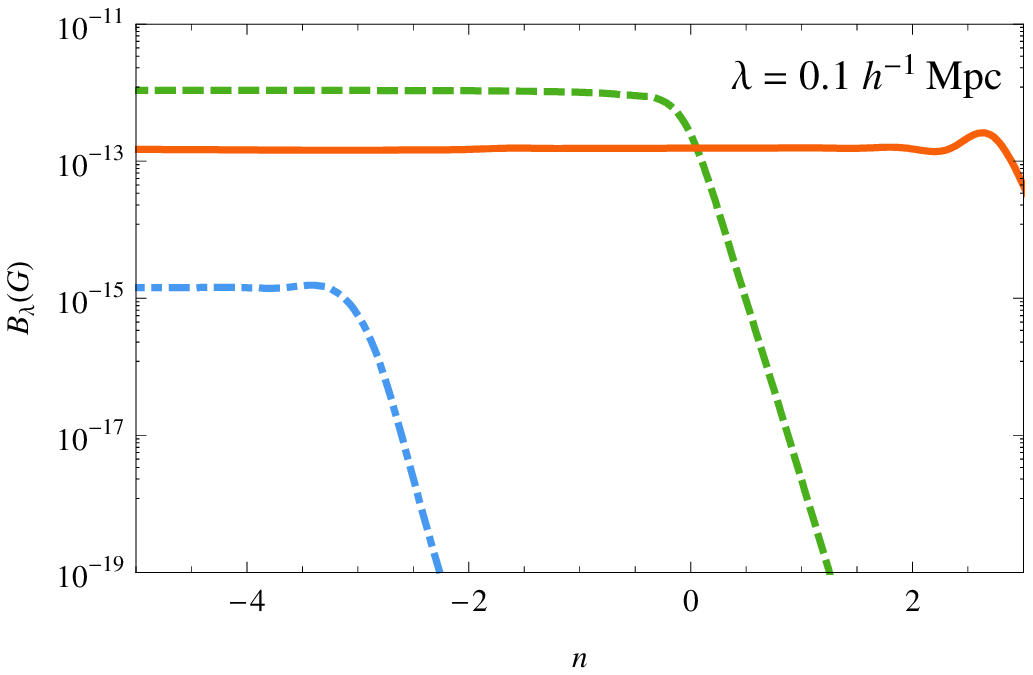}}
{\epsfxsize=8.5cm\epsfbox{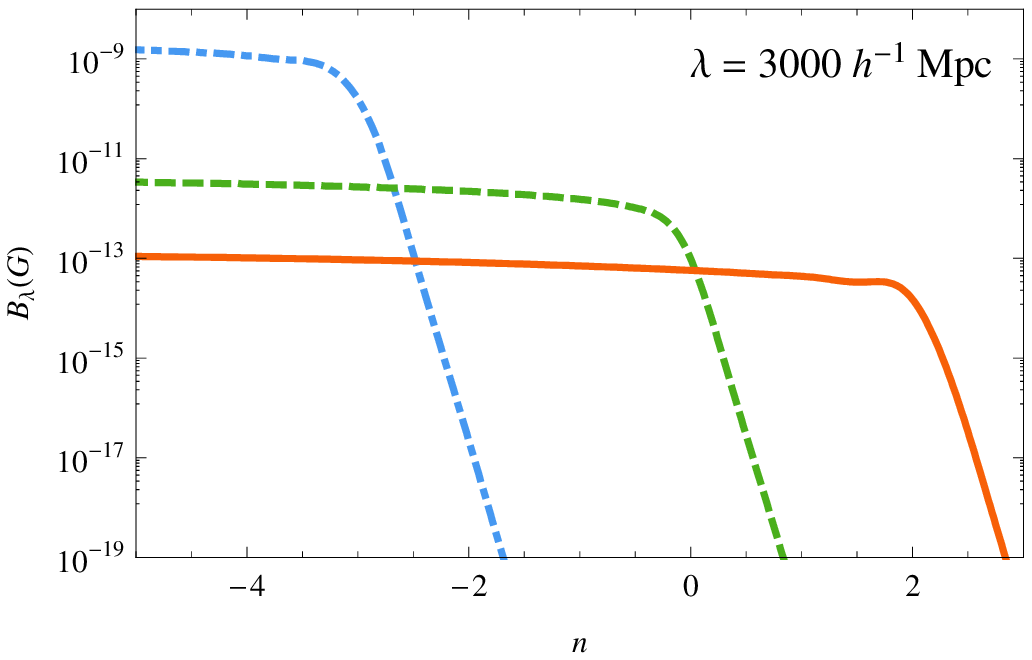}} 
\vspace*{-0.3cm}
\caption{ Lower limits on the magnetic
fields generated  on galactic scales (upper panel)
and Hubble horizon scales (lower panel) in terms
of the magnetic spectral index $n$ for different values
of the vorticity spectral index $m$. Dot-dashed blue for $m=0$, dashed 
green
for $m\simeq -3$ and full red for $m\simeq -5$.}
\end{center}
\end{figure}

Vorticity perturbations generate anisotropies in the 
CMB temperature at recombination time whose amplitude 
should be compatible with 
present observations.  Taking into 
account the scaling properties of vorticity derived before, 
such limits on 
a scale $\lambda$ can be written today as \cite{FC}:
\begin{eqnarray}
\omega_\lambda^2\lsim  \frac{l^2C_l/(2\pi)z_{rec}^2G(\lambda,m)}
{8l^3(l+1)R(l,m)}
\label{ol}
\end{eqnarray}
where $l^2C_l/(2\pi)\simeq 10^{-10}$ and 
\begin{eqnarray}
R(l,m)=\int_{k_{rec}}^{\infty}dk 
\frac{j_l^2(k\eta_*)}{(k\eta_*)^2}k^m
\end{eqnarray}
with $\eta_*=\eta_0-\eta_{rec}$. We will consider the minimum of 
the r.h.s. of (\ref{ol}) with respect to $l$ which, for $m<-1$, 
is located at $l\sim 29$
 and, for $m>-1$, at $l\sim 1200$  which is the highest 
multipole measured by WMAP.

These stringent upper limits on vorticity  
can be translated  using (\ref{fund}) into lower limits
on the present value of the  magnetic field created by the
effective current. 
For the sake of concreteness, 
we  take $H_{k0}\simeq 2 \times 10^{-6}$ eV in  (\ref{chPE}).  This  
value corresponds to a  scale of inflation around $100$ GeV i.e. 
in the electroweak range. It  
satisfies $H_{k0}\lsim M_\Lambda$ and also the limits on
the primordial electromagnetic fluctuations coming from 
their imprint on CMB anisotropies (see \cite{EM3}). 
We have evaluated numerically the 
 integrals appearing in (\ref{fund}) for $\epsilon\simeq 0.01$, although  
the $\epsilon$-dependence of the bounds is very small. 
In Fig. 1 we show  the 
lower limits on the magnetic fields
generated by this mechanism on scales $\lambda=0.1 h^{-1}$ Mpc,  
(which is the relevant scale for  
galaxies and clusters (see \cite{CMBDurrer})),
and  $\lambda=3000 h^{-1}$ Mpc.
 These results show that the produced fields could
have strong amplitudes even in the largest scales 
and act as seeds for a galactic
dynamo or even play the role of primordial fields 
and account for observations just by amplification 
due to the collapse and differential rotation of the protogalactic cloud.

It is interesting
to note that, since super-Hubble modes of 
the effective electromagnetic current 
are not generated, we expect magnetic fields to be present only on 
sub-Hubble scales. This means, that the constraints coming
from the dissipation of super-Hubble magnetic fields
into  gravity waves  before nucleosynthesis \cite{NucDurrer}
do not apply in the present case.  In any case, these results
show that a more precise 
determination of the magnetic and vorticity spectra
on cosmological scales could help 
establishing the feasibility of the extended theory in (\ref{action})
for producing the observed cosmic magnetic
fields.

\vspace*{0.09cm}

{\em Acknowledgments:} We would like to thank Ruth Durrer and Misao Sasaki
for useful comments. 
This work has been  supported by
MICINN (Spain) project numbers
FIS 2008-01323 and FPA
2008-00592, CAM/UCM 910309, 
MEC grant BES-2006-12059 and MICINN Consolider-Ingenio 
MULTIDARK CSD2009-00064. 
J.B. also received support from the Norwegian Research Council under the YGGDRASIL
project no 195761/V11 and wishes to thank the hospitality of the University 
of Geneva where part of this work was performed.


\vspace*{-0.65cm}





\begin{thebibliography}{0}
\bibitem{Widrow}  L.~M.~Widrow,
  Rev.\ Mod.\ Phys.\  {\bf 74} (2002) 775; R.~M.~Kulsrud and E.~G.~Zweibel,
  Rept.\ Prog.\ Phys.\  {\bf 71} (2008) 0046091; P.~P.~Kronberg,
  Rept.\ Prog.\ Phys.\  {\bf 57} (1994) 325.
\bibitem{extragalactic} A. Neronov and I. Vovk, 
Science {\bf 328} (2010) 73; F.~Tavecchio, et al.,  
arXiv:1004.1329 [astro-ph.CO];
S.~'i.~Ando, A.~Kusenko, Astrophys.\ J.\  {\bf 722 } (2010)  L39;
 A.~Neronov, D.~V.~Semikoz, P.~G.~Tinyakov {\it et al.}, 
[arXiv:1006.0164 [astro-ph.HE]].
\bibitem{Brandenburg} A. Brandenburg and K. Subramanian, 
Phys. Rept. {\bf 417} (2005) 1
\bibitem{young}M.~L.~Bernet, F.~Miniati, S.~J.~Lilly, P.~P.~Kronberg and M.~Dessauges-Zavadsky,
  Nature {\bf 454} (2008) 302;
  A.~M.~Wolfe, R.~A.~Jorgenson, T.~Robishaw, C.~Heiles and J.~X.~Prochaska,
  Nature {\bf 455} (2008) 638
\bibitem{Harrison} E.R. Harrison, MNRAS {\bf 147} (1970) 279; 
Phys. Rev. Lett. {\bf 30} (1973) 18
\bibitem{Turner} M.S. Turner and L.M. Widrow, {\it Phys. Rev.} {\bf D37}:
2743, (1988); B. Ratra, Astrophys. J. {\bf 391} (1992) L1; 
K.~Bamba and J.~Yokoyama,  Phys.\ Rev.\  D {\bf 69} (2004) 043507;
 K.~Bamba and M.~Sasaki,  JCAP {\bf 0702} (2007) 030

\bibitem{QLS}J.~M.~Quashnock, A.~Loeb and D.~N.~Spergel, 
Astrophys. J.  {\bf 344} (1989) L49; T. Vachaspati, Phys. Lett. {\bf B265}
(1991) 258
\bibitem{pert} O.~Bertolami and D.~F.~Mota,
  Phys.\ Lett.\  B {\bf 455} (1999) 96; A.~L.~Maroto, 
  Phys.\ Rev.\  D {\bf 64} (2001) 083006; K.~Ichiki, et al. 
Science {\bf 311} (2006) 827

\bibitem{EM1} J. Beltr\'an Jim\'enez and A.L. Maroto, 
JCAP {\bf 0903} (2009) 016; Int.\ J.\ Mod.\ Phys.\  D {\bf 18} (2009) 2243
\bibitem{EM2} J. Beltr\'an Jim\'enez and A.L. Maroto, 
Phys.\ Lett.\  B {\bf 686} (2010) 175
\bibitem{EM3} J.~B.~Jimenez, T.~S.~Koivisto, A.~L.~Maroto and D.~F.~Mota,
  JCAP {\bf 0910} (2009) 029
\bibitem{Itzykson} C. Itzykson and J.B. Zuber, {\it Quantum Field Theory},
McGraw-Hill (1980); N.N. Bogoliubov and D.V. Shirkov, {\it
Introduction to the theory of quantized fields}, Interscience
Publishers, Inc. (1959).
\bibitem{Dolgov} A.~Dolgov and J.~Silk,
  Phys.\ Rev.\  D {\bf 47} (1993) 3144.
\bibitem{FC} C.~Caprini and P.~G.~Ferreira,
  JCAP {\bf 0502} (2005) 006
\bibitem{Tanaka}
  Y.~Urakawa and T.~Tanaka,
  Prog.\ Theor.\ Phys.\  {\bf 122} (2009) 779
\bibitem{Durrer} R. Durrer, {\it The Cosmic Microwave Background}, 
Cambridge (2008)
\bibitem{CMBDurrer}
  R.~Durrer, P.~G.~Ferreira and T.~Kahniashvili,
  Phys.\ Rev.\  D {\bf 61} (2000) 043001
\bibitem{Barrow} K. Subramanian and J.D. Barrow, Phys. Rev. D {\bf 58} (1998)
083502
\bibitem{NucDurrer}
  C.~Caprini and R.~Durrer,
  Phys.\ Rev.\  D {\bf 65} (2001) 023517; 
C.~Caprini, R.~Durrer, E.~Fenu,  JCAP {\bf 0911 } (2009)  001;
T.~Kahniashvili, A.~G.~Tevzadze, S.~K.~Sethi {\it et al.},
  Phys.\ Rev.\  {\bf D82 } (2010)  083005;
 T.~Kahniashvili, A.~G.~Tevzadze, B.~Ratra, [arXiv:0907.0197 [astro-ph.CO]].













\end{thebibliography}
\end{document}